# Phase-dependent Supermode Excitation in Photonic Molecules

**Yury E. Geints**

V.E. Zuev Institute of Atmospheric Optics SB RAS, 1 Zuev square, 634021 Tomsk, Russia
ygeints@iao.ru

**Abstract**

A photonic molecule (PM) is a miniature diffractive optical structure composed of resonance microcavities called atoms (e.g., cylinders or spheres) supporting a set of high-quality eigenmodes. All atoms in a PM are coupled by the electromagnetic fields of eigenmodes, which form collective supermodes of the whole PM. We consider a particular type of mirror-symmetric PMs being optically excited simultaneously via two light channels (tapered fibers). Based on the numerical simulations, we show that the spectral composition of supermodes in such PM can be effectively manipulated by changing the phase detuning between the optical channels. For a seven-atom silicon microcylinder cyclic-PM is demonstrated the possibility to achieve tenfold intensity amplification/suppression of several supermodes from the Stokes and anti-Stokes bands of PM spectrum.

## 1. Introduction

Recently, specific photonic structures comprised by the electromagnetic coupling of two or more nearby located optical microcavities called the "photonic molecules" (PM) have gained increasing scientific interest [Bayer1998, Rackovich2010, Liao2020]. This molecular analogy stems from the specific physical properties of PM manifested in the fact that the normal optical modes of a PM consisting of several interacting "photonic atoms" and the electronic states of real atoms of a matter combined into a molecule behave in a similar way. In other words, a single-atom optical resonator is treated as a "photonic atom", and several optically coupled resonators are treated as a photonic analogue of a real molecule. With these higher-dimensional types of composite photonic structures, a controlled interaction between light and matter can be achieved and enhanced by manipulating the coupling or matching of individual cavities including their mechanical and optical tuning. In this respect, PM is the subject of research in a more general scientific field known as the topological photonics [Ozawa2019]. The development and study of PM not only adds new functionality to the design of optical devices based on microresonators, but also paves the way for their application to the study of simulation techniques in atomic physics and quantum optics [Xia2007, Liao2020, Hoang2022].

As elementary resonance cavities, i.e. the photonic atoms, photonic nanostructures supporting standing-wave eigenmodes in the spatial type of the whispering gallery modes (WGMs) are most commonly used. However, there are several PM implementations on 2D photonic crystals [Bose2013], coupled quantum dots [Wang2022], multi-mode optical fibers [Kalinin2021], and plasmonic nanocavities [Wonmi2012]. Dielectric WGM resonators can also have different sizes and geometric shapes such as spheres, disks, toroids, or cylinders [Mikaiyama1999,

Smotrova2006, Gagnon2014]. Here, a common practice is the requirement of a sufficiently high-quality factor of the microcavity to provide WGM generation. Since the surface of a dielectric particle is not a perfect reflector, the optical fields of the eigenmodes leak out the boundaries of the particle and form an evanescent mantle background around a resonator. In this sense, the WGMs in a dielectric particle are the quasi-normal electromagnetic modes. Optical coupling of two or more atoms in the PM is precisely due to these evanescent fields [Mikaiyama1999, Kanaev2006].

To date, quite a few different topological forms of PMs have been studied ranging from basic diatomic molecules and linear 1D-chains of many atoms composed of several closely spaced resonance cavities, to complex 2D-structures in the form of quasi-closely packed particle micro-assemblies with cyclic symmetry or without it [PhotMicro]. Meanwhile, a common property of all PMs is the appearance of specific collective electromagnetic oscillations in the microresonator lattice, called the superlattice modes or supermodes (SMs), which result from the collective resonance interaction of all photonic atoms in the PM [Evans1996]. According to the coupled modes theory (CMT) [Liao2020], when a pair of resonators is physically close to each other their optical eigenmodes begin to interact and produce a new hybridized eigenmode spectrum of the cavity pair as a whole [Li2017]. Similarly, the hybridization of two atomic orbitals in a hydrogen molecule takes place with the appearance of new spectral lines. Thus, two initially isolated resonators exhibiting different (or even the same) eigenmodes produce new set of modes with certain eigenfrequencies and phases when combined in a PM because of the electromagnetic coupling between the photonic atoms.

Like real molecule electron orbitals, the supermodes of PMs differ in the type of optical atom coupling and form bounding (even) and anti-bounding (odd) modes, which means the corresponding symmetry or antisymmetry of the electric field distribution relative to the PM axis [Bayer1998, PhotMicro]. For a PM formed by identical atoms with a resonance frequency of $\omega_0$, bounding SMs have the eigenfrequencies in the Stokes spectral band, $\omega_b < \omega_0$, while the anti-bounding SMs are conversely blue-shifted, $\omega_a > \omega_0$. Besides, with a change of the type of inter-atom coupling, the SM resonance quality factor ($Q$) also changes demonstrating in some cases a dramatic increase relative to the value of an isolated resonator (in a certain PM topology and at a certain space between atoms) [Boriskina2006]. Fano-resonances excitation [Kanaev2006, Hoang2017], as well as diffraction coupling of optical modes after eigenvalue degeneracy removing [Huang2019] can be considered as the probable reasons for such decrease of the optical losses in supermodes.

Normally, the supermodes are excited using a single optical channel located at one side of a PM and built either as a critical coupled tapered fiber [Li2017] or free-space coupling radiation originated from a laser or an electric dipole [Hoang2022]. Both types of SM generation imply the transfer of optical excitation from a photonic atom, or several photonic atoms, directly illuminated by an optical radiation, sequentially to their distant neighbors through the interaction of resonance mode fields (direct coupling [Li2010]). Although this method is undoubtedly simple, its disadvantage is the relatively low excitation efficiency of all resonant cavities in a PM due to high diffraction losses when propagating the input optical excitation between atoms.

In the present work, we consider another situation, when the excitation of superresonant states of a spatially symmetric PM is carried out simultaneously from two opposite sides similarly to, e.g., a dissipative sensor based on WGM resonant cavity placed inside a conical capillary [Ren2019, Jiang2022]. In contrast to the previous studies, the photonic structure considered uses two single-mode waveguides (fibers) operating at the same optical frequency, but with different input phases to feed the input optical excitation to the PM. Changing the phase delay between the excitation channels makes it possible to amplify or suppress the generation of certain PM supermodes and thus control its spectral response. To our knowledge, this is the first (although theoretical) demonstration of SM manipulating in a photonic molecule without changing its internal structure (the inter-atomic gap). Obviously, by attributing to the phase delay of optical channels a certain physical origin, e.g., fiber density fluctuations, temperature/pressure load of one of the waveguides, one can principally propose a design of miniature sensor of the corresponding physical stimulus.

## 2. PM Simulation Model

Hereinafter, two particular PM topologies are considered constituted from identical cylinder-shaped atoms located in the nodes of rectangular or hexagonal 2D-spatial lattice (Fig. 1a), which will be further abbreviated as (R)-PM and (H)-PM, respectively. PM is assembled on a silicon oxide ($SiO_2$) substrate and is structurally characterized by the atomic diameter ($d$) and inter-atomic gap ($g$). Crystalline silicon (Si) is chosen as the photonic atom material. Two waveguides made of optical fiber are fabricated on the PM sides at some distance from the outermost atoms. The fibers support propagation of a single $TE_{00}$ mode (Fig. 1b), which has an electric field vector directed along the $z$-axis. The optical radiation with the wave number $k = 2\pi/\lambda$ ($\lambda$ is the wavelength) can be fed into one of the waveguides (say, channel 1) with a certain phase delay $\varphi$. The spectral band of input radiation is chosen sufficiently broad to cover a given WGM resonance of the PM atom. Thus, in the case under study, the intrinsic $TE_{7,1}$ resonance with the quality-factor $Q_0 = 7800$ is chosen which is excited inside a silicon cylinder with the diameter $d = 1.2$ µm at the wavelength $\lambda_0 = 1.3383$ µm. The refractive index of silicon $n = 3.5$ with near-zero optical absorption ($\sim 10^{-11}$) is adopted in the near-infrared wavelength band [Schinke2015].

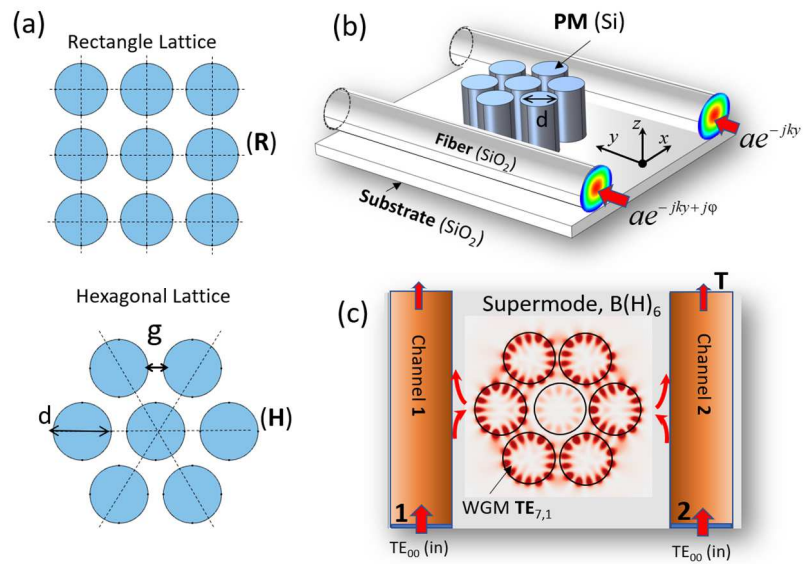

Fig. 1. (a) Axisymmetric PM types: rectangle (R) and hexagonal (H) spatial lattices. (b) Computational 3D-model of the photonic structure considered with a (H)-PM and two fiber optical waveguides pumped with an optical phase difference $\varphi$. (c) COMSOL 2D-model of a (H)-PM.

For the calculation of the optical fields in a PM, we use the "Wave Optics" module of COMSOL Multiphysics 5.1 computer software, which numerically solves the wave equation for full-vectorial electromagnetic field using the finite elements method (FEM). Without limiting the generality of the consideration, we reduce the problem dimension and exploit the 2D-model of PM shown in Fig. 1(c). The whole photonic structure comprised of a PM and two fiber waveguides is surrounded by a rectangular region of perfectly matched (absorbing) layers (PML) to prevent multiple wave reflections from the domain boundaries. Optical radiation is input through digital ports 1 and 2 located on the substrate bottom edge. All domains are discretized using a numerical mesh with triangular elements and maximum edge size $\lambda/40$. Wavelength sweep is performed by creating a parametric study within $1.333\ \mu m \leq \lambda \leq 1.343\ \mu m$ range, which includes all SMs of the considered PM topologies based on $TE_{7,1}$ resonance of the isolated atom.

## 3. Coupled Mode Analysis of PM Supermodes

Before proceeding to complex PM structures, it is instructive to analyze the coupled oscillations in the simplest system of two closely spaced identical resonant cavities pumped through two optical channels, as shown in Fig. 2(a). Assume, each cavity in the pair supports a single degenerate optical eigenmode with the wavelength $\lambda_0$.

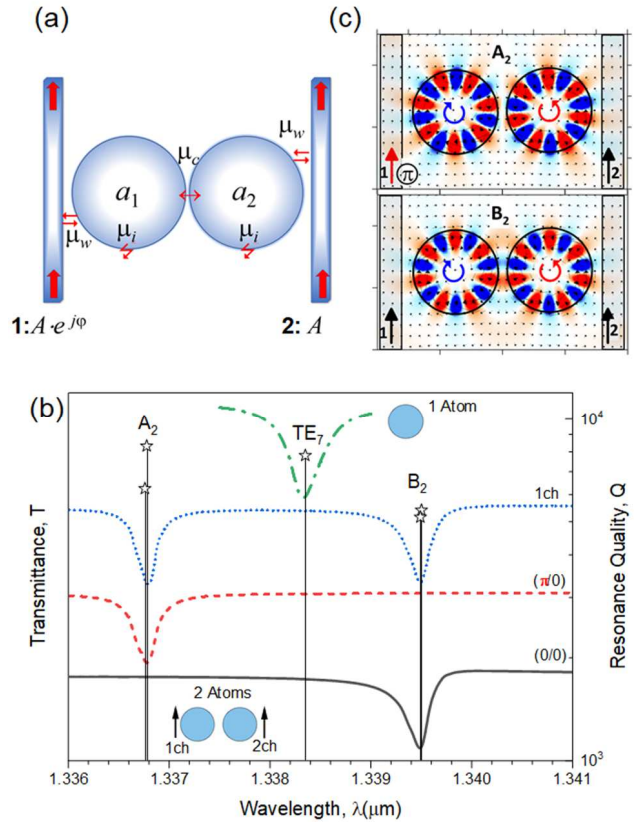

Fig. 2. (a) Schematic of two coupled resonant cavities excited through two optical channels. (b) Spectral transmission ($T$) of the channel 2 as a part of a silicon two-atom PM in the dependence on phase difference between the optical channels denoted as (1ch/2ch). (c) Electric field amplitude ($E_z$) and energy flux (Poynting vector) spatial distributions near cylindrical atoms in the resonance conditions of two PM supermodes, $A_2$ (anti-bonding) and $B_2$ (bonding).

According to the CMT [Manolatou1999, Li2010], the evolution in time (*t*) of two eigenmodes $a_1$ and $a_2$ existing in a cavity pair is described by the following equations:

$$\frac{d}{dt}a_1 = \left[j\omega_0 - (\mu_i + \mu_w)\right]a_1 + A\sqrt{\mu_w}e^{j\varphi} - j\mu_c a_2$$
$$\frac{d}{dt}a_2 = \left[j\omega_0 - (\mu_i + \mu_w)\right]a_2 + A\sqrt{\mu_w} - j\mu_c a_1 \quad (1)$$

Here, $\omega_0 = 2\pi c/\lambda_0$ is the cyclic frequency of resonant mode excited in a single cavity, $\mu_i = \omega_0/2Q$ is field amplitude decrement due to intrinsic losses in the cavity (absorption, radiative escape, etc.), $\mu_w$, $\mu_c$ stay for the normalized coupling parameters of resonators with a waveguide and with each other, respectively. Obviously, the inter-cavity coupling of the fields will depend on the size of the gap *g*, i.e. $\mu_c = \mu_c(g)$. In the following we take $A = 1$, and then the solution to Eq. (1) for any of the normal modes, e.g., $a_1$, can be written as follows:

$$a_1(t) \propto e^{-\mu_t t} \left\{\left[j(\mu_c - \omega_0) + \mu_t\right]\left(e^{j\varphi} - 1\right)e^{j(\omega_0 + \mu_c)t} - \left[j(\mu_c + \omega_0) - \mu_t\right]\left(e^{j\varphi} + 1\right)e^{j(\omega_0 - \mu_c)t}\right\} \quad (2)$$

where $\mu_t = (\mu_i + \mu_w)$.

This expression shows that a pair of coupled resonators instead of a single resonant mode $\omega_0$ is already characterized by two collective eigenvalues – the supermodes, with the eigenfrequencies $\omega_{1,2} = \omega_0 \pm \mu_c$, equidistantly shifted from the resonant frequency of the isolated cavity by the value of the coupling coefficient $\mu_c$ and having the power ratio: $a_1^2/a_2^2 \approx (\mu_c - \omega_0)^2/(\mu_c + \omega_0)^2$. At the same time, the magnitude of the phase shift φ between the excitation channels of the photonic duet controls the amplitudes of these supermodes. In two extreme cases, $\varphi = 2\pi m$ and $\varphi = (2m+1)\pi$, where *m* is an integer, the anti-bounding or bounding modes are completely suppressed, respectively.

This can be clearly seen in Fig. 2(b), which plots the spectrum of the transmission function *T* (in power) of the second excitation channel of the two-atom PM under consideration at different phase delays $\varphi = 0$ and $\pi$. The gap *g* between the cylindrical atoms is set to 300 nm. This figure shows, that if $\varphi = \pi$, the short-wavelength anti-bonding SM ($A_2$) with $\lambda_1 = 1.3368$ µm is resonantly excited, and this mode is not excited at all if $\varphi = 0$. For the symmetric bounding supermode ($B_2$) in the red spectral wing with $\lambda_2 = 1.3395$ µm, in contrast, the exact opposite situation is observed. For comparison, in the absence of optical radiation in channel 1, the spectral behavior of the transmission *T* (shown by blue dashed line) exhibits two minima corresponding to both supermodes with approximately equal amplitudes.

This result can be obtained from Eq. (2), if one set the phase shift of excitation channels equal to some intermediate value, say, $\varphi = \pi/2$. Interestingly, both supermodes are twice degenerate in frequency but with different quality factors *Q*. The quality factor of the supermode resonances is also shown by asterisks in Fig. 2(b). As seen, all the anti-bonding SMs ($A_2$) generally have higher quality factors, and one of them has $Q = 8300$ that is higher than the quality factor of the $TE_{7,1}$ WGM excited in isolated cylindrical atom.

## 4. Phase-Controlled Supermodes in a PM

The spatial distribution of electric field amplitude $E_z$ of two SMs is shown in Fig. 2(c). According to the SM notation, in the anti-bonding SM the phases of optical oscillations in both atoms are opposite and mirror-symmetric with respect to the PM symmetry axis. In fact, the anti-bonding mode can be regarded as the eigenmode of the half-PM placed near the perfectly electric conducting plane (PEC mirror). As shown in Ref. [Liu2000] on the example with a dielectric sphere, the blue shift of TE-mode resonance is realized in this case, what is evident for the $A_2$ supermode also. Bounding supermode $B_2$ of PM demonstrates the appearance of strong bonds between the electric fields of atoms. As in an ordinary molecule of matter, in a PM the interpenetration and hybridization of eigenmodes (energy levels) take place, which is expressed in the appearance of new low-frequency collective optical oscillations (low-energy molecular sublevels).

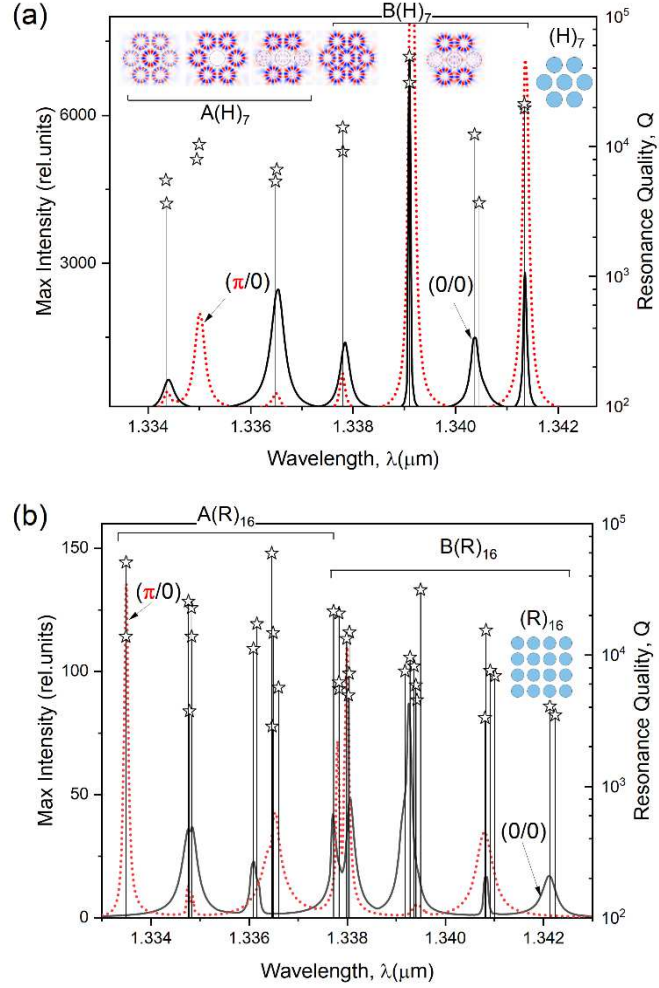

Fig. 3. (a, b) PM spectral response in (a) hexagonal and (b) rectangular topologies depending on the phase delay φ between the optical channels denoted as (1ch/2ch).

Noteworthy, regardless of the type of electromagnetic coupling of atoms in PM, the optical energy fluxes in both supermodes are similar. This follows from the Poynting vector maps, $\mathbf{S} = (c/8\pi)\,\mathrm{Re}\!\left[\mathbf{E}\times\mathbf{H}^*\right]$ (where $\mathbf{E}$ and $\mathbf{H}$ are the electric and magnetic vectors, respectively, $c$ is the light speed), which are plotted by arrows in Fig. 2(c). As is readily observed, both atoms in both SM configurations always have counter-circulation of optical energy within the WGM volumes without formation of any regions of singular energy flow, i.e., optical vortices or saddles, which normally occur in the standing wave resonant cavity [Geints2022-OAO].

An increase in the number of atoms in PM leads to a significant complication in the structure of its spectral fingerprint due to the appearance of new SM configurations. Figures 3(a,b) show the spectral responce of two PM comprised of seven $(H)_7$ and sixteen $(R)_{16}$ atoms in hexagonal and rectangular topologies with inter-atom gap $g = 300$ nm. The maximum value of the squared electric field, $I_m = \{|E_z|^2\}_{PM}$, achieved inside the atoms is plotted on the vertical axis.

As seen, in the case of molecule $(H)_7$ its spectrum contains a total of 14 collective supermodes, but each of them is twice degenerate. The spatial configurations of three anti-bonding $A(H)_7$ and two bonding $B(H)_7$ supermodes are shown in Fig. 3(a) by the pictograms. The spectrum of molecule $(R)_{16}$ contains about 10 supermodes with two- and threefold frequency degeneracy. Importantly, the excitation efficiency of some of these SMs depends dramatically on the phase delay between the optical channels pumping the photonic molecule. Most clearly this is manifested, e.g., for the anti-bonding mode with $\lambda = 1.335$ μm and the bonding SM at $\lambda = 1.3404$ μm in molecule $(H)_7$ having circular topology, and also for analogous modes at $\lambda = 1.3335$ μm and $1.3421$ μm in molecule $(R)_{16}$ with the rectangular atom configuration.

According to Eq. (2), in a chosen supermode a gradual increase/decrease of the intensity $I_{A/B}$ takes place by changing the phase $\varphi$ in one of the optical channels exciting the PM. This tendency follows the relationship: $I_{A/B}(\varphi) \propto (1 \mp \cos\varphi)$, whereas the SM intensity ratio is given by the square tangent of the half-phase: $I_A/I_B \propto tg^2(\varphi/2)$. This tendency is shown in Fig. 4, where the intensities of anti-bonding ($I_A$) and bonding ($I_B$) cyclic supermodes of molecule $(H)_7$ and their ratio $I_A/I_B$ are plotted as a function of $\varphi$. Evidently, the relative change of intensities $f_\pi = I(\varphi=0)/I(\varphi=\pi)$ of either supermode in the two extreme cases can exceed an order of magnitude. Meanwhile, the ratio of SM intensities at both frequencies ($I_A/I_B$) exhibits already more than 200-fold change in its value when varying the phase in the range from 0 to $\pi$ (Inset in Fig. 4).

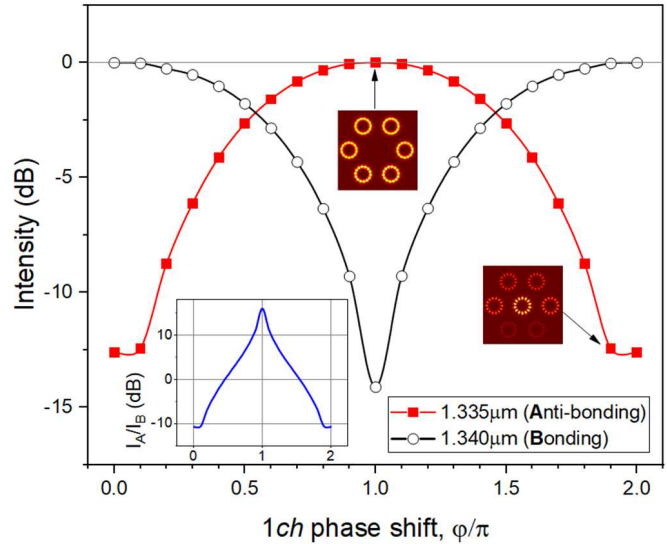

Fig. 4. Phase-control of PM supermodes. Relative intensity of anti-bonding and bonding supermodes in $(H)_7$ PM in the dependence on the phase delay between optical channels. Inset: Logarithm of anti- and bonding supermode intensity ratio.

Worthwhile noting, when the optical excitation phase of one of the waveguides is switched between 0 and $\pi$, not only the intensity, but also the optical field distribution of SM changes dramatically. Actually, phase switching causes the degeneracy removal of the corresponding eigenmode by damping of the most intensive eigenoscillation. Thus, for the above chosen anti-bonding resonance $A(H)_7$ (1.335 µm) at the phase delay $\varphi = \pi$, the supermode with the highest intensity is realized in the spatial form of cyclically excited atoms, as shown by the pictogram in Fig. 4. Switching the excitation phase to another extreme case, $\varphi = 0$, leads to destructive interference of WGM fields in the left and right parts of PM, because their relative phases should be antisymmetric in this supermode. As a result, the optical field in the annular atoms of PM sharply decreases the intensity against the resonance of the central atom, the excitation of which does not depend on symmetry conditions.

Compared with the WGM in a single atom governed by the only two key parameters being the microcylinder radius and its refractive index, the PM supermodes have one more degree of freedom as the interatomic gap, which leads to the appearance of a *g*-dependence of the spectral position and quality factor of each SM. This beneficial SM property is commonly used in the design of optofluidic sensors [PhotMic, Mandal2008] and coupled-cavity microlasers [Smotrova2006, Wang2022] when the superimposed deformations of the photonic atom lattice allow to regulate the excitation efficiency of supermodes.

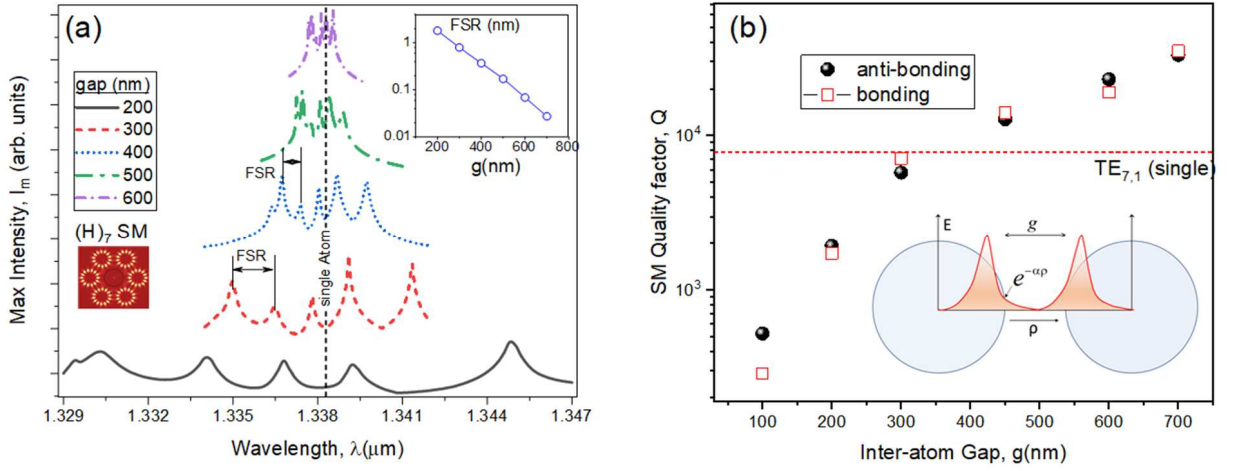

Fig. 5. (a) SM spectrum of $(H)_7$ photonic molecule at interatomic gap variation. Inset: Supermode FSR versus *g*. (b) Quality factor *Q* of two SMs in the Stokes and anti-Stokes spectrum wings of PM in the dependence on the inter-atom spacing.

This is illustrated in Figs. 5(a) and (b), which show the spectrum of supermodes of a hexagonal $(H)_7$-PM with different spacing between atoms, as well as the quality factor change of two circular-type SMs located at the edges of the Stokes and anti-Stokes regions of the molecule spectrum (see Fig. 4). Here, we also plot the dependence of the averaged spectral distance between neighboring supermodes (FSR) when the interatomic gap *g* changes. It follows from the figures that bringing the atoms in PM closer together monotonically reduces the spectral range occupied by all the supermodes which are associated with the chosen WGM resonance of the isolated atom (in this case $TE_{7,1}$). As the interval *g* increases, the mode eigenfrequencies get closer and their resonant profiles begin to overlap, which eventually leads to an increase in the effective quality factor of each SM, mainly due to the appearance of Fano resonances [Hoang2017]. As can be seen in Fig. 5(b), the quality factor of both collective modes composed of cyclically excited atoms, increases with increasing the interparticle spacing, since this reduces the overlap of the WGM fields of neighboring atoms and decreases the associated radiation losses from the resonant cavities.

Indeed, the asymptotic representation for the electric field $E$ of a WGM leaking out a spherical dielectric resonant cavity is as follows [Little1999]: $E(\rho) \propto \exp(-\alpha\rho)$, and represents an evanescent tail with the damping $\alpha = \sqrt{l(l+1)-(kr)^2}$, where the distance $\rho$ is normalized to the cavity radius $r$ and is measured from the cavity rim (see the diagram in Fig. 5b). It is easy to obtain that for the considered photonic atom parameters ($r = 600$ nm, $l = 7$, $\lambda = 1.3$ μm) the distance $\rho^*$ giving a $1/e^2$ drop in the field amplitude is of the order of $\rho^* = 2/\alpha \approx 180$ nm. This means that the fields coupling of the neighboring resonant cavities (atoms) can be considered insignificant if they are at a distance $g = 2\rho^* \approx 360$ nm. In Fig. 5(b) one can see that approximately at this distance the quality factor $Q$ of collective supermode oscillations of all PM atoms becomes close in value to the quality factor of $TE_{7,1}$ WGM of isolated cylinder atom ($Q = 7800$). Denser atom placement in the molecule will inevitably increase the overlap of the mode fields and decrease the effective quality factor of the supermode. Particularly, this does not allow the building of the so-called super-resonator [Yariv1999] on the base of PM constructed from close-contacting looped atoms [Hoang2017].

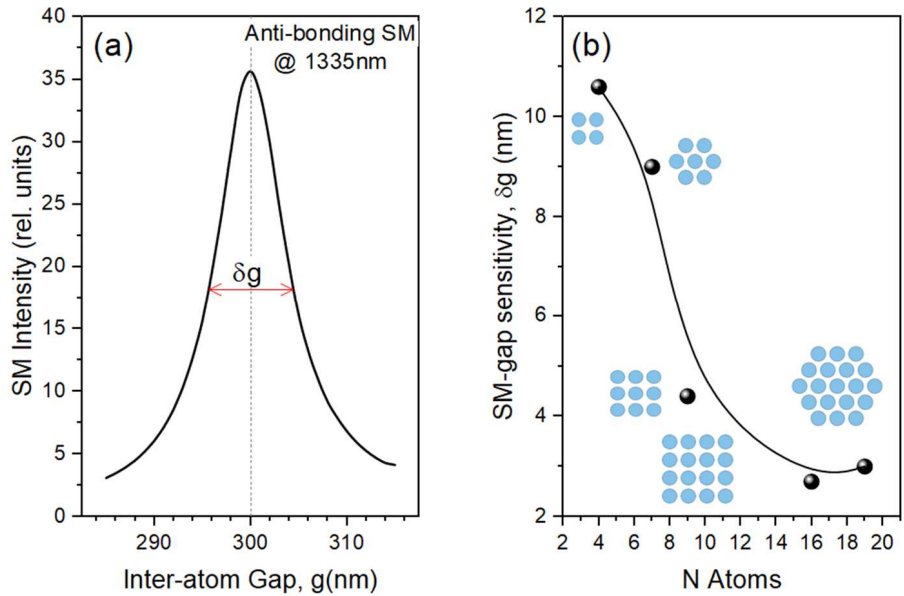

Fig. 6. SM response to the interatomic gap variation. (a) A(H)$_7$ SM intensity in the dependence on $g$. (b) SM gap-sensitivity $\delta g$ for different PM topologies.

The sensitivity of PM supermodes to the changes in the interatomic gap is illustrated by the graphs in Fig. 6(a,b). Here, one of the anti-bonding PM modes is exemplified and its intensity is calculated by varying the distance between the atoms about some prefixed value ($g = 300$ nm). As seen from Fig. 6(a), the functional dependence of SM excitation efficiency on the parameter $g$ exhibits a typical resonance profile with a certain half-width $\delta g$, which, in turn, depends on PM structural type and the number of atoms (Fig. 6b). For the simplest 4-atom symmetric PM, the SM sensitivity parameter is $\delta g \approx 11$ nm and decreases with increasing complexity of the molecular structure, reaching about 3 nm for the 16-atom (R)-PM. Further increase in the atomic number does not lead to a significant change in $\delta g$ value. Worthwhile noting, the results presented are obtained at a specific "unperturbed" value of the gap $g$. As our numerical simulations show, choosing a different interatomic distance, greater or less than 300 nm, changes the absolute values of δ$g$ but does not influence the very trend of the relationship presented.

The limiting value of the SM intensity ratio $f_\pi$ at the extreme phase shifts between the optical excitation channels, φ = 0 and φ = π, also depends significantly on the topology of a particular supermode and the PM structure itself. The $f_\pi$ dynamics for the anti-bounding SMs in different PMs is shown in Figs. 7(a,b) when the gap between the atoms changes. Clearly, there is a certain range of the interatomic gap when the ratio of SM intensities becomes maximum. Further approaching or removal of atoms leads to a sharp drop in $f_\pi$.

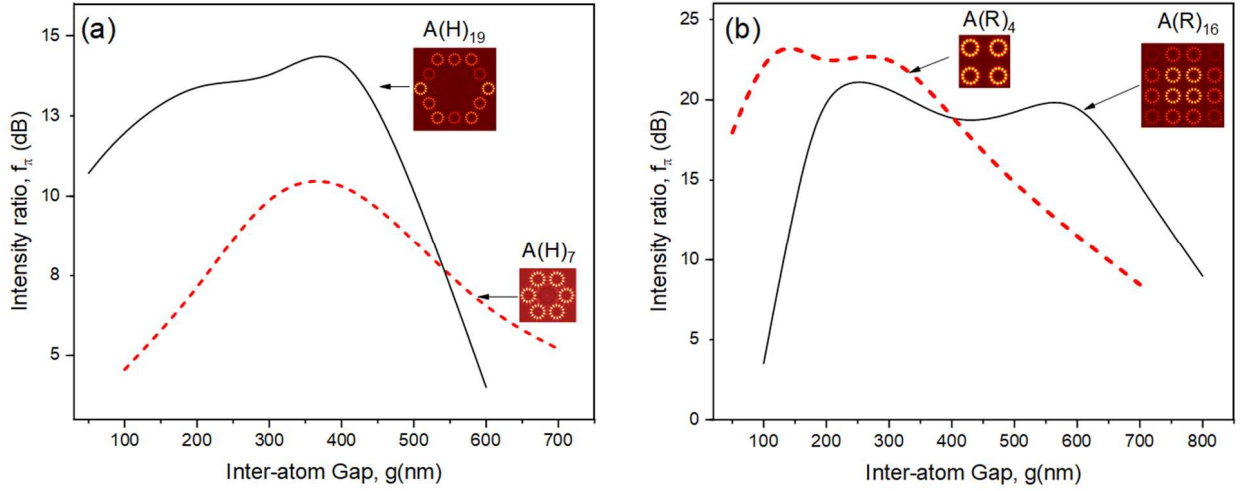

Fig. 7. Limiting intensity ratio $f_\pi$ of anti-bonding SMs at interatomic gap variation in (a) hexagonal (H) and (b) rectangular (R) molecule topology.

This optimal range of interatomic spacing increases with the complexity of the molecular structure. For example, for a PM with a rectangular topology (Fig. 7b), a 4-atom molecule has a plateau in $f_\pi(g)$ dependence in the range from 100 nm to 270 nm. At the same time, the 16-atom molecule shows similar behavior of $f_\pi$ parameter in a wider (although right-shifted) range from $g$ = 200 nm to 600 nm.

The ratio of polar-dephased SM intensities in molecules with a hexagonal atom topology is generally lower than in (R)-PM and more sensitive to the packing density of atoms. As follows from Fig. 7(a), for such PM the behavior of $f_\pi(g)$ dependence demonstrates not a plateau but rather a clearly pronounced maximum at $g ≈ 400$ nm. Interestingly, at this value of the gap the quality factor $Q$ of collective atom oscillations becomes close in value to the quality factor of $TE_{7,1}$ WGM of an isolated atom (shown by red dashed line in the inset to Fig. 5b). The physical background to this correlation is not yet clear and demands further investigation.

## 5. Conclusion

To conclude, using FEM based numerical simulations, the collective optical field resonances (supermodes) are studied in a certain class of photonic molecules with mirror symmetry of photonic atoms made of silicon microcylinders and pumped by optical radiation. Specific feature of the photonic structures under study is the presence of two channels of optical excitation organized as two optical microfibers on both sides of the photonic molecule. We show that by adjusting the phase delay between the channels it is possible to effectively control the molecule supermode spectrum causing dumping or amplifying different types of supermodes within broad limits.

High sensitivity of the supermode intensity to the phase of the optical excitation allows one to propose a sensor design based on the photonic molecule for detection and measuring various kinds of physical stimuli on the optical waveguide feeding the sensor which impact its dielectric constant (mechanical load, medium humidity, temperature). Meanwhile, structurally it is not necessary to use two optical channels for the excitation of photonic molecule; it is quite possible to loop the same fiber providing the required phase delay to excite or suppress the chosen supermode(s). In this design, the optical excitation is fed from opposite corners of the photon molecule, and the fiber loop section acts as the sensing element, where the measuring wave phase progression arises upon an external stimulus.